\documentclass{article}[11pt]
\usepackage{amsfonts}
\usepackage{CJK,amsmath,indentfirst}
\usepackage{graphics}
\usepackage{graphicx}
\begin{document}

\title{\bf Interactions Between Solitons and Other Nonlinear Schr\"odinger Waves}

\author{\footnotesize S. Y. Lou$^{1,2,3}$\thanks{Email: lousenyue@nbu.edu.cn}, Xue-Ping Cheng$^{2}$, Chun-li Chen$^{4}$ and Xiao-Yan Tang$^{2}$\\
\footnotesize $^{1}$ \it Shanghai Key Laboratory of Trustworthy Computing, East China Normal University, Shanghai 200062, China\\
\footnotesize $^{2}$\it Department of Physics, Shanghai Jiao Tong University, Shanghai, 200240, China\\
\footnotesize $^{3}$\it Faculty of Science, Ningbo University, Ningbo, 315211, China\\
\footnotesize $^{4}$\it Department of Mathematics, Shanghai Jiao Tong University, Shanghai, 200240, China
}
\date{}
\maketitle
\parindent=0pt
\textbf{Abstract:} The Nonlinear Schr\"odinger (NLS) equation is widely used in everywhere of natural science. Various nonlinear excitations of the NLS equation have been found by many methods. However, except for the soliton-soliton interactions, it is very difficult to find interaction solutions between different types of nonlinear excitations. In this paper, three very simple and powerful methods, the symmetry reduction method, the truncated Painlev\'e analysis and the generalized tanh function expansion approach, are further developed to find interaction solutions between solitons and other types of NLS waves. Especially, the soliton-cnoidal wave interaction solutions are explicitly studied in terms of the Jacobi elliptic functions and the third type of incomplete elliptic integrals.
In addition to the new method and new solutions of the NLS equation, the results can unearth some new physics. The solitons may be decelerated/accelerated through the interactions of soliton with background waves which may be utilized to study tsunami waves and fiber soliton communications;
the static/moving optical lattices may be automatically excited in all mediums described by the NLS systems; solitons elastically interact with non-soliton background waves, and the elastic interaction property with only phase shifts provides a new mechanism to produce a controllable routing switch that is applicable in optical information and optical communications.

\vskip.4in
\renewcommand{\thesection}{\arabic{section}}
\parindent=20pt
\section{Introduction}
The soliton and/or solitary wave equations connect rich histories of
exactly solvable systems constructed in mathematical, statistical and many-body physics, and powerfully demonstrate the unity of nonlinear concepts across disciplines
and scales from micro-physics and biology to cosmology \cite{KdV}.  Among these equations, the nonlinear Schr\"odinger (NLS) equation
\begin{equation}
p_t+\frac12ibp_{xx}-i|p|^2p=0,\quad i\equiv \sqrt{-1},\quad b\equiv \pm1,\label{NLS}
\end{equation}
is the most ubiquitous \cite{NLSE}. Originally, the NLS equation is derived to describe the envelope dynamics of a quasi-monochromatic plane wave propagating in a weakly nonlinear dispersive medium when dissipative processes are negligible (see for instance \cite{FL}). The NLS equation finds an important application in plasma physics, where it describes
electron (Langmuir) waves \cite{PL}. The NLS equation in nonlinear optics is also well known
to describe
self-modulation and self-focusing of light in a Kerr-type
nonlinear medium \cite{OP}. The great current interest in the NLS application is initiated by the prediction of the solitons in nonlinear optical fibers \cite{FIB}
and the concept of the soliton laser \cite{LAS}.
Furthermore, the NLS equation is widely used in ferromagnets with easy-axis anisotropy, molecular chains, nonideal Bose gas, nuclear matter, solid state
medium, gravity waves, optical lattice, Bose-Einstein condensations, and so on \cite{SOON}.

The multiple soliton solutions of the NLS equation have been obtained by many authors via different methods, say, the Hirota's bilinear method \cite{Hirota}, the Darboux transformations (DT) \cite{DT} and the B\"acklund transformations (BT) \cite{BT}. Using DT and BT, in principle, one can obtain a new solution from a known one. However, in practice, one can only find multiple soliton solutions stemming from simple constant solutions. It is rather difficult to find new explicit solutions starting from nonconstant nonlinear waves such as the cnoidal waves and Painlev\'e waves via DT and BT. In Refs. \cite{C-S,cs1}, the mutisoliton complexes on a cnoidal wave background have been studied by the DTs for the multi-component NLS equations, the sine-Gordon (SG) equation and the Toda lattice.

Recently, it is found that combining the symmetry reduction method and the DT or BT related nonlocal symmeries\cite{Nonlocal}, one can readily find the interaction solutions among solitons and other nonlinear excitations including the cnoidal waves for the KdV \cite{n1} and KP \cite{n2} equations. In this paper, much simpler while more effective methods are developed for the NLS system while they are also valid for any other integrable systems.

In real physics, to decelerate and/or accelerate soliton is one of the important topics. For instance, the propagation of the tsunami wave can be described by a soliton (or a solitary wave) \cite{tsunami}, then to find a possible mechanism to decelerate tsunami waves is definitely significant. In optics, to decelerate and/or to accelerate solitons also plays an important role \cite{Bragg}. In this paper, we propose a new physical mechanism to accelerate/decelerate optical solitons via soliton interactions with background waves. In optics information and optics communications, to produce a controllable routing switch is also an important task. Below, we will present a new possible mechanism to solve this problem.

In section 2 of this paper, we review the known local symmetries at first and the investigate the novel symmetries, the nonlocal symmetries for the Ablowitz-Kaup-Newell-Segur (AKNS) system.
In section 3, to find the finite transformation related to a special nonlocal symmetry, the nonlocal symmetry for the original AKNS system is localized for an enlarged AKNS system. Thus the finite auto-B\"acklund transformation (ABT) theorem is naturally obtained by Lie's first principle.
In section 4, thanks to the localization procedure of the last section, the group invariant solutions related to the nonlocal symmetries of the AKNS system are obtained by means of the symmetry reduction method.
In section 5, hinted by the results of Sec. 4, two much simpler but more general methods, the truncated Painlev\'e expansion and the hyperbolic tangent function expansion method are \bf {\em extended} \rm to get interaction solutions among different types of nonlinear waves.
 Some special exact solutions and their possible new physics are discussed in Sec. 6. The last section is a summary and discussion.

\section{Infinitely many local and nonlocal symmetries of the AKNS system}

It is known that for an integrable system, there exist infinitely many symmetries. Using symmetries of a nonlinear system, one can obtain various interesting results of the model especially to study its exact solutions of nonlinear systems \cite{Olver} via symmetry reduction method. However, one usually uses the local symmetries to find symmetry reductions while the existence of infinitely many nonlocal symmetries \cite{inv} is ignored by most of scientists. Recently, we find that the nonlocal symmetries can be successfully used to find some types of important interaction solutions which are difficult to be found by other approaches \cite{Nonlocal,n1}.

For simplicity and generality, we consider the NLS equation as a special case of the AKNS system
\begin{subequations}\label{AKNS}
\begin{equation}
p_t+\frac12ibp_{xx}-ip^2q=0,
\end{equation}
\begin{equation}
q_t-\frac12ibq_{xx}+iq^2p=0,
\end{equation}
\end{subequations}
with $q$ being a conjugate of $p$, i.e.,
$
q=p^*.
$

A symmetry, $\sigma$,
\begin{equation}\label{sym}
\sigma\equiv \left(\begin{array}{c}\sigma^p \cr \sigma^q\end{array}\right)
\end{equation}
is defined as a solution of the linearized equation of the AKNS equation
\begin{subequations}\label{pqsym}
\begin{equation}
\sigma_{t}^p+\frac12ib\sigma_{xx}^p-2ipq\sigma^p-ip^2\sigma^q=0,
\end{equation}
\begin{equation}
\sigma_{t}^q-\frac12ib\sigma_{xx}^q+2ipq\sigma^q+iq^2\sigma^p=0,
\end{equation}
\end{subequations}
 which means the AKNS system \eqref{AKNS} is form invariant under the transformation
\begin{equation}\label{sym1}
\left(\begin{array}{c}p \cr q\end{array}\right)\rightarrow \left(\begin{array}{c}p \cr q\end{array}\right)+\epsilon \left(\begin{array}{c}\sigma^p \cr \sigma^q\end{array}\right)
\end{equation}
with infinitesimal parameter $\epsilon$.

The infinitely many local $K$-symmetries
\begin{equation}\label{symk}
K_n=\Phi^n \left(\begin{array}{c}-ip \cr iq\end{array}\right)
\end{equation}
and $\tau$-symmetries
\begin{equation}\label{symt}
\tau_{n+1}=\Phi^n \tau_1,\quad \tau_1=\left(\begin{array}{c} tp_x+ibxp \cr tq_x-ibxq\end{array}\right),\ n=0,\ 1,\ 2,\ \cdots
\end{equation}
are known in literature \cite{KT}. In \eqref{symk} and \eqref{symt}, the recursion operator $\Phi$ is defined as
\begin{equation}\label{symt}
\Phi=\left(\begin{array}{cc}-\partial+2bp\partial^{-1}q & 2bp\partial^{-1}p \cr -2bq\partial^{-1}q & \partial-2bq\partial^{-1}p \end{array}\right)
\end{equation}
with $\partial=\partial/\partial_x,\ \partial^{-1}=\int_{-\infty}^xdx$.
The $K_n$ symmetries constitute a commute symmetry algebra and the $\tau_n$ symmetries constitute the centerless Virasoro symmetry algebra. Especially, the symmetries
$$K_0=\left(\begin{array}{c}-ip \cr iq\end{array}\right),\
K_1=\left(\begin{array}{c}p_x \cr q_x\end{array}\right),\
K_2=-2ib\left(\begin{array}{c}p_t \cr q_t\end{array}\right),
$$ and $$\tau_0=\left(\begin{array}{c} tp_x+ibxp \cr tq_x-ibxq\end{array}\right),\ \tau_1=-ib\left(\begin{array}{c} 2tp_t+(xp)_x \cr 2tq_t+(xq)_x\end{array}\right)$$
constitute a finite dimensional Lie point symmetry algebra with nonzero commutators
\begin{eqnarray}\label{alg}
[\tau_0,\ K_1]=i b K_0,\  [\tau_0,\ K_2]=2ibK_1,\
[\tau_1,\ K_1]=i b K_1,\ [\tau_1,\ K_2]=2ibK_2,\
[\tau_1,\ \tau_0]=-i b\tau_0,
\end{eqnarray}
where the commutator $[A,\ B]$ of
$$A=\left(\begin{array}{c} A_1 \cr A_2\end{array}\right),\ B=\left(\begin{array}{c} B_1 \cr B_2\end{array}\right)$$
is defined as
\begin{eqnarray}\label{com}
[A,\ B]=A'B-B'A
\end{eqnarray}
with
\begin{eqnarray}\label{com}
A'=\left(\begin{array}{cc} A'_{1p} & A'_{1q} \cr A'_{2p} & A'_{2q}\end{array}\right)
\end{eqnarray}
and
$$ A'_{1p} f \equiv \left.\frac{\partial}{\partial
\epsilon} A_1(p+\epsilon f)\right|_{\epsilon=0}.$$

To find infinitely many nonlocal symmetries, one can use some different approaches. For instance the inverse recursion operator method \cite{inv}, the infinitesimal forms of the Darboux transformations \cite{DT} or B\"acklund transformations \cite{BT}, the conformal invariance of the Schwarzian forms \cite{con}, the derivatives of the inner parameters \cite{DT} and the higher order Lax operators \cite{NKP} or the infinitely many nonhomogeneous Lax pairs \cite{DT}.

For the AKNS system \eqref{AKNS}, its Lax pair possesses the following form \cite{AS}
\begin{equation}\label{Lx}
\left(\begin{array}{c}\phi_1 \cr \phi_2\end{array}\right)_x=\left(\begin{array}{cc}-i\lambda & \frac{p}{\sqrt{b}} \cr \frac{q}{\sqrt{b}} & i\lambda \end{array}\right)\left(\begin{array}{c}\phi_1 \cr \phi_2\end{array}\right),
\end{equation}
\begin{equation}\label{Lt}
\left(\begin{array}{c}\phi_1 \cr \phi_2\end{array}\right)_t=\left(\begin{array}{cc}ib\lambda^2+\frac12ipq & -\frac12ip_x-p\lambda \cr \frac12iq_x-q\lambda & -ib\lambda^2-\frac12ipq \end{array}\right)\left(\begin{array}{c}\phi_1 \cr \phi_2\end{array}\right).
\end{equation}
A simple nonlocal symmetry of the AKNS system related to the Lax pair is the so-called square eigenfunction symmetry \cite{NAKNS}
\begin{equation}\label{s0}
N_0=\left(\begin{array}{c}\phi_1^2 \cr \phi_2^2\end{array}\right).
\end{equation}
Then the infinitely many nonlocal symmetries can be written as
\begin{equation}\label{sn}
N_n=\Phi^{-n}N_0, \quad n=0,\ 1,\ 2,\ \cdots,\ \infty,
\end{equation}
where the inverse recursion operator has been given in \cite{invAKNS}.

Alternatively, because the AKNS system \eqref{AKNS} is $\lambda$ independent while the nonlocal symmetry $N_0$ is $\lambda$ dependent, we can obtain infinitely many nonlocal symmetries via parameter $\lambda$ derivatives of $N_0$:
\begin{equation}\label{sn}
N_n=\frac1{n!}\frac{\mbox{d}^n}{\mbox{d} \lambda^n}N_0, \quad n=0,\ 1,\ 2,\ \cdots,\ \infty.
\end{equation}
To write down $N_n$ of \eqref{sn} explicitly, we can introduce infinitely many nonhomogeneous Lax pairs with the help of the original Lax pair \eqref{Lx} and \eqref{Lt},
\begin{equation}\label{Lxn}
\left(\begin{array}{c}\phi_{1,n} \cr \phi_{2,n}\end{array}\right)_x=\left(\begin{array}{cc}-i\lambda & \frac{p}{\sqrt{b}} \cr \frac{q}{\sqrt{b}} & i\lambda \end{array}\right)\left(\begin{array}{c}\phi_{1,n} \cr \phi_{2,n}\end{array}\right)+\left(\begin{array}{cc}-i & 0 \cr 0 & i \end{array}\right)\left(\begin{array}{c}\phi_{1,n-1} \cr \phi_{2,n-1}\end{array}\right),
\end{equation}
\begin{eqnarray}\label{Ltn}
\left(\begin{array}{c}\phi_{1,n} \cr \phi_{2,n}\end{array}\right)_t&=&\left(\begin{array}{cc}ib\lambda^2+\frac12ipq & -\frac12ip_x-p\lambda \cr \frac12iq_x-q\lambda & -ib\lambda^2-\frac12ipq \end{array}\right)\left(\begin{array}{c}\phi_{1,n} \cr \phi_{2,n}\end{array}\right)\nonumber\\
&& +\left(\begin{array}{cc}2ib\lambda & -p \cr -q & -2ib\lambda \end{array}\right)\left(\begin{array}{c}\phi_{1,n-1} \cr \phi_{2,n-1}\end{array}\right)+\left(\begin{array}{cc}ib& 0 \cr 0 & -ib \end{array}\right)\left(\begin{array}{c}\phi_{1,n-2} \cr \phi_{2,n-2}\end{array}\right)
\end{eqnarray}
for $\ n=0,\ 1,\ 2,\ \cdots,$ and $\phi_{1,n<0}=\phi_{2,n<0}=0$. It is not difficult to verify that all the consistent conditions $(\phi_{j,n})_{xt}=(\phi_{j,n})_{tx}$ for all $j=1,\ 2,\ n=0,\ 1,\ 2,\ \cdots$ are just the AKNS system \eqref{AKNS}.

With the help of the Lax pairs \eqref{Lxn} and \eqref{Ltn}, the infinitely many nonlocal symmetries can be simply expressed as
\begin{equation}\label{isn}
N_n=\sum_{k=0}^n\left(\begin{array}{c}\phi_{1,k}\phi_{1,n-k} \cr \phi_{2,k}\phi_{2,n-k}\end{array}\right),\ \quad n=0,\ 1,\ 2,\ \cdots.
\end{equation}
Furthermore, the arbitrariness of the spectral parameter $\lambda$ also makes it possible to the existence of infinitely many square eigenfunction symmetries
\begin{equation}\label{bsn}
\bar{N}_n=\left(\begin{array}{c}\phi_{1n}^2 \cr \phi_{2n}^2\end{array}\right),
\end{equation}
with
\begin{subequations}\label{Lnxt}
\begin{equation}
\left(\begin{array}{c}\phi_{1n} \cr \phi_{2n}\end{array}\right)_x=\left(\begin{array}{cc}-i\lambda_n & \frac{p}{\sqrt{b}} \cr \frac{q}{\sqrt{b}} & i\lambda_n \end{array}\right)\left(\begin{array}{c}\phi_{1n} \cr \phi_{2n}\end{array}\right),
\end{equation}
\begin{eqnarray}\label{Lntn}
\left(\begin{array}{c}\phi_{1n} \cr \phi_{2n}\end{array}\right)_t&=&\left(\begin{array}{cc}
ib\lambda_n^2+\frac12ipq & -\frac12ip_x-p\lambda_n \cr \frac12iq_x-q\lambda_n & -ib\lambda_n^2-\frac12ipq \end{array}\right)\left(\begin{array}{c}\phi_{1n} \cr \phi_{2n}\end{array}\right).
\end{eqnarray}
\end{subequations}

\section{Localization of nonlocal symmetries}
Now, one of the important question is what kind of finite transformations are related to the nonlocal symmetries. In this section, we only concentrate on the finite transformation of the square eigenfunction symmetry $N_0$ \eqref{s0}. According to the Lie's first principle, to find the finite transformation of $N_0$, one has to solve the ``initial value" problem
\begin{subequations}\label{ivp}
\begin{equation}
\frac{\mbox{d} p'(\epsilon)}{\mbox{d} \epsilon}=\phi'^2_{1}(\epsilon),\ p'(0)=p,\
\end{equation}
\begin{equation}
\frac{\mbox{d} q'(\epsilon)}{\mbox{d} \epsilon}=\phi'^2_{2}(\epsilon),\ q'(0)=q,
\end{equation}
\end{subequations}
with
\begin{subequations}\label{Lxta}
\begin{equation}
\left(\begin{array}{c}\phi'_{1}(\epsilon) \cr \phi'_{2}(\epsilon)\end{array}\right)_x=\left(\begin{array}{cc}-i\lambda & \frac{p'(\epsilon)}{\sqrt{b}} \cr \frac{q'(\epsilon)}{\sqrt{b}} & i\lambda \end{array}\right)\left(\begin{array}{c}\phi'_{1}(\epsilon) \cr \phi'_{2}(\epsilon)\end{array}\right),
\end{equation}
\begin{eqnarray}
\left(\begin{array}{c}\phi'_{1}(\epsilon) \cr \phi'_{2}(\epsilon)\end{array}\right)_t&=&\left(\begin{array}{cc}
ib\lambda^2+\frac12ip'(\epsilon)q'(\epsilon) & -\frac12ip'_x(\epsilon)-p'(\epsilon)\lambda \cr \frac12iq'_x(\epsilon)-q'(\epsilon)\lambda & -ib\lambda_n^2-\frac12ip'(\epsilon)q'(\epsilon) \end{array}\right)\left(\begin{array}{c}\phi'_{1}(\epsilon) \cr \phi'_{2}(\epsilon)\end{array}\right).
\end{eqnarray}
\end{subequations}
Because of the presence of $\phi'_{1}(a)$ and $\phi'_{2}(a)$ in the initial value problem \eqref{ivp}, we have to study the corresponding symmetry transformation for the spectral functions $\phi_1$ and $\phi_2$ related to the symmetry $N_0$ for $p$ and $q$. In other words, we have to solve the symmetry equations
\begin{eqnarray}\label{s34x}
\left(\begin{array}{c}\sigma^{\phi_{1}}_x \cr \sigma^{\phi_{2}}_x \end{array}\right)&=&\left(\begin{array}{cc}-i\lambda & \frac{p}{\sqrt{b}} \cr \frac{q}{\sqrt{b}} & i\lambda \end{array}\right)\left(\begin{array}{c}\sigma^{\phi_{1}} \cr \sigma^{\phi_{2}}\end{array}\right)+\left(\begin{array}{cc}\frac{\phi_2}{\sqrt{b}} & 0 \cr 0 & \frac{\phi_1}{\sqrt{b}} \end{array}\right)\left(\begin{array}{c}\sigma^{p} \cr \sigma^{q}\end{array}\right) + i\sigma^{\lambda}\left(\begin{array}{c}-\phi_1 \cr \phi_2\end{array}\right),
\end{eqnarray}
and
\begin{eqnarray}\label{s34t}
\left(\begin{array}{c}\sigma^{\phi_{1}}_t \cr \sigma^{\phi_{2}}_t \end{array}\right)&=&\left(\begin{array}{cc}ib\lambda^2+\frac12ipq & -\frac12\sqrt{b}(ip_x+2p\lambda) \cr \frac12\sqrt{b}(iq_x-2q\lambda) & -ib\lambda^2-\frac12ipq \end{array}\right)\left(\begin{array}{c}\sigma^{\phi_{1}} \cr \sigma^{\phi_{2}}\end{array}\right)+\sqrt{b}\sigma^\lambda
\left(\begin{array}{c}2i\sqrt{b}\lambda\phi_1-p\phi_2 \cr -2i\sqrt{b}\lambda\phi_2-q\phi_1\end{array}\right)\nonumber\\
&&
+\left(\begin{array}{cc}
\frac{i}2q\phi_1-\frac{\sqrt{b}}2\phi_2(2\lambda+i\partial_x) & \frac{i}2p\phi_1 \cr -\frac{i}2q\phi_2 & -\frac{i}2p\phi_2+\frac{\sqrt{b}}2\phi_1(i\partial_x -2\lambda)
 \end{array}\right)\left(\begin{array}{c}\sigma^{p} \cr \sigma^{q}\end{array}\right),
\end{eqnarray}
with
\begin{equation}\label{spq}
\sigma^{p}=\phi_1^2,\quad \sigma^{q}=\phi_2^2.
\end{equation}
It is not difficult to verify that Eqs. \eqref{s34x}, \eqref{s34t} and \eqref{spq} possess the solution
\begin{equation}\label{sp12}
\sigma^{\phi_1}=\phi\phi_1,\quad \sigma^{\phi2}=\phi\phi_2,
\end{equation}
with
\begin{subequations}\label{fxt}
\begin{equation}\label{fx}
\phi_x=\frac1{\sqrt{b}}\phi_1\phi_2,
\end{equation}
\begin{equation}\label{ft}
\phi_t=\frac{i}2(q\phi_1^2-p\phi_2^2)-2\sqrt{b}\lambda\phi_1\phi_2.
\end{equation}
\end{subequations}
Similarly, due to the entrance of $\phi$ in \eqref{sp12}, we have to study the solution of the symmetry equation for the field $\phi$
\begin{subequations}\label{sfxt}
\begin{equation}\label{sfx}
\sigma^\phi_x=\frac1{\sqrt{b}}(\sigma^{\phi_1}\phi_2+\sigma^{\phi_2}\phi_1),
\end{equation}
\begin{equation}\label{sft}
\sigma^\phi_t=\frac{i}2(\sigma^q\phi_1^2-\sigma^p\phi_2^2
+2q\phi_1\sigma^{\phi_1}-2p\phi_2\sigma^{\phi_2})
-2\sqrt{b}\lambda(\sigma^{\phi_1}\phi_2+\sigma^{\phi_2}\phi_1),
\end{equation}
\end{subequations}
with \eqref{spq} and \eqref{sp12}. One can prove that the consistent condition of \eqref{sfxt}, $\phi_{xt}=\phi_{tx}$, is identically satisfied.

It is straightforward to find that the meaningful solution of \eqref{sfxt} with \eqref{spq} and \eqref{sp12} has the form
\begin{equation}\label{sf}
\sigma^\phi=\phi^2.
\end{equation}

Thus, the nonlocal symmetry $N_0$ of the AKNS system is localized for the prolonged system \eqref{AKNS}, \eqref{Lx}, \eqref{Lt} and \eqref{fxt}. Meanwhile, the initial value problem \eqref{ivp} is changed to
\begin{subequations}\label{ivpf}
\begin{equation}
\frac{\mbox{d} p'(\epsilon)}{\mbox{d} \epsilon}=\phi'^2_{1}(\epsilon),\ p'(0)=p,\
\end{equation}
\begin{equation}
\frac{\mbox{d} q'(\epsilon)}{\mbox{d} \epsilon}=\phi'^2_{2}(\epsilon),\ q'(0)=q,
\end{equation}
\begin{equation}
\frac{\mbox{d} \phi'_1(\epsilon)}{\mbox{d} \epsilon}=\phi'(\epsilon)\phi'_1(\epsilon),\ \phi'_1(0)=\phi_1,
\end{equation}
\begin{equation}
\frac{\mbox{d} \phi'_2(\epsilon)}{\mbox{d} \epsilon}=\phi'(\epsilon)\phi'_2(\epsilon),\ \phi'_2(0)=\phi_2,
\end{equation}
\begin{equation}
\frac{\mbox{d} \phi'(\epsilon)}{\mbox{d} \epsilon}=\phi'^2(\epsilon),\ \phi'(0)=\phi.
\end{equation}
\end{subequations}
After solving out the initial value problem \eqref{ivpf}, we have the following auto-B\"acklund transformation (ABT) theorem:\\
\bf ABT theorem. \rm If $\{p,\ q,\ \phi_1,\ \phi_2,\ \phi\}$ is a solution of the prolonged AKNS system \eqref{AKNS}, \eqref{Lx}, \eqref{Lt} and \eqref{fxt}, so is $\{p'(\epsilon),\ q'(\epsilon),\ \phi'_1(\epsilon),\ \phi'_2(\epsilon),\ \phi'(\epsilon)\}$ with
\begin{subequations}\label{ABT}
\begin{equation}
p'(\epsilon)=p+\frac{\epsilon\phi_1^2}{1-\epsilon \phi},
\end{equation}
\begin{equation}
q'(\epsilon)=q+\frac{\epsilon\phi_2^2}{1-\epsilon \phi},
\end{equation}
\begin{equation}
\phi'_1(\epsilon)=\frac{\phi_1}{1-\epsilon \phi},
\end{equation}
\begin{equation}
\phi'_2(\epsilon)=\frac{\phi_2}{1-\epsilon \phi},
\end{equation}
\begin{equation}\label{MT}
\phi'(\epsilon)=\frac{\phi}{1-\epsilon \phi}.
\end{equation}
\end{subequations}

\section{Symmetry reductions of the AKNS system with nonlocal symmetries}
Symmetry reduction is one of the powerful method to study exact solutions of nonlinear systems \cite{Olver}. However, one usually uses the local symmetries to find symmetry reductions while the existence of infinitely many nonlocal symmetries \cite{inv} is ignored by most of scientists. Recently, we find that the nonlocal symmetries can be successfully used to obtain some types of important interaction solutions which are difficult to be found by other approaches \cite{Nonlocal,n1}.

In this section, we study the symmetry reductions of the AKNS system \eqref{AKNS} under the local symmetries $\{K_0,\ K_1,\ K_2,\ \tau_0,\ \tau_1\}$ and the nonlocal symmetry $N_0$ which is corresponding to the infinitesimal form of the B\"aclund transformation. As in the KdV case \cite{n1}, to find symmetry  reductions related to the nonlocal symmetry, we have to enlarge the original system such that the nonlocal
symmetry can be localized for the enlarged system.

For the AKNS system, its enlarged system has been given in the last section. It is easy to demonstrate that the general Lie point symmetry solution of the enlarged AKNS system \eqref{AKNS}, \eqref{Lx}, \eqref{Lt} and \eqref{fxt} has the form
\begin{equation}\label{symp}
\sigma_{nl}=\left(\begin{array}{c}
\sigma^p\\
\sigma^q\\
\sigma^{\phi_1}\\
\sigma^{\phi_2}\\
\sigma^\phi\\
\sigma^\lambda
\end{array}\right)
=\left(\begin{array}{c}
(c_5x+c_4t+c_2)p_x+(2c_5t+c_3)p_t+(-c_1+c_5+c_4bix)p+c_6\phi_1^2\\ (c_5x+c_4t+c_2)q_x+(2c_5t+c_3)q_t+(c_1+c_5-c_4bix)q+c_6\phi_2^2\\ (c_5x+c_4t+c_2)\phi_{1x}+(2c_5t+c_3)\phi_{1t}+
(c_1(a-\frac12)+c_6\phi+\frac12c_4bix)\phi_1\\ (c_5x+c_4t+c_2)\phi_{2x}+(2c_5t+c_3)\phi_{2t}
+(c_1(a+\frac12)+c_6\phi-\frac12c_4bix)\phi_2\\ (c_5x+c_4t+c_2)\phi_{1x}+(2c_5t+c_3)\phi_{1t}+(2c_1a-c_5+c_6\phi)\phi\\ c_5\lambda-\frac{b}2c_4\end{array}\right).
\end{equation}
The last component of the symmetry \eqref{symp} implies that for the enlarged system the scaling and Galileo invariance must company by the transformation of the spectral parameter $\lambda$.

To find the symmetry reductions related to the nonlocal symmetry, i.e., to find group invariant solutions related to the symmetry \eqref{symp} with $c_6\neq0$, two nontrivial cases should be considered for $c_5\neq0$ and $c_5=0$.\\
\bf Case 1 $c_5\neq0$. \rm In this case, the group invariant solution can be solved from the invariant condition $\sigma_{nl}=0$. The final result has the form
\begin{subequations}\label{red1}
\begin{equation}
\phi=\frac{c_5C_1}{c_6}\big[\tanh\big(F(\xi)+C_1\ln(\tau)\big)\big],\ \xi=\frac{x+c_5}\tau+\frac{c_4\tau}{2c_5^2},\ \tau=\sqrt{2c_5t+c_3},
\end{equation}
\begin{equation}
\phi_1=\Phi_1(\xi)\mbox{sech}\big(F(\xi)+C_1\ln(\tau)\big)\tau^{\frac1{4a}
[2a(ibc_4C_2-1)+2C_1+1]}\exp\left[-\frac{ibc_4\tau}{c_5^3}
(c_4\tau+4c_5^2\xi)\right],
\end{equation}
\begin{equation}
\phi_2=\Phi_2(\xi)\mbox{sech}\big(F(\xi)+C_1\ln(\tau)\big)\tau^{-\frac1{4a}
[2a(ibc_4C_2+1)+2C_1+1]}\exp\left[\frac{ibc_4\tau}{c_5^3}
(c_4\tau+4c_5^2\xi)\right],
\end{equation}
\begin{equation}
p=\left[P(\xi)-\frac{c_6}{c_5C_1}\Phi^2_1(\xi)
\tanh\big(F(\xi)+C_1\ln(\tau)\big)\right]\tau^{
ibc_4C_2-1+\frac{2C_1+1}{2a}}\exp\left[-\frac{ibc_4\tau}{4c_5^3}
(c_4\tau+4c_5^2\xi)\right],
\end{equation}
\begin{equation}
q=\left[Q(\xi)-\frac{c_6}{c_5C_1}\Phi^2_2(\xi)\tanh
\big(F(\xi)+C_1\ln(\tau)\big)\right]\tau^{
-ibc_4C_2-1-\frac{2C_1+1}{2a}}\exp\left[\frac{ibc_4\tau}{4c_5^3}
(c_4\tau+4c_5^2\xi)\right],
\end{equation}
\end{subequations}
where $C_1\equiv \frac{ac_1}{c_5}-\frac12,\ C_2\equiv\frac{c_2}{c_5^2}-\frac{c_3c_4}{2c_5^3}$, while the group invariant functions $F(\xi),\ P(\xi),\ Q(\xi),\ \Phi_1(\xi)$ and $\Phi_2(\xi)$ should satisfy symmetry reduction equations which can be obtained by substituting \eqref{red1} into the prolonged system \eqref{AKNS}, \eqref{Lx}, \eqref{Lt} and \eqref{fxt}. It is straightforward to find the final reduction equations
\begin{subequations}\label{redeq1}
\begin{equation}
F_\xi=\frac{c_6\Phi_1\Phi_2}{c_5C_1\sqrt{b}},
\end{equation}
\begin{equation}
Q=\frac{2ic_5\xi\Phi_2}{\sqrt{b}\Phi_1}+\frac{P\Phi_2^2}{\Phi_1^2}
-\frac{2ic_5^2C_1^2}{c_6\Phi_1^2},
\end{equation}
\begin{equation}
\Phi_{1\xi}=\frac{P\Phi_2}{\sqrt{b}},
\end{equation}
\begin{equation}
\Phi_{2\xi}=\frac{Q\Phi_1}{\sqrt{b}},
\end{equation}
\begin{equation}
P_\xi=\frac{P^2\Phi_2}{\sqrt{b}\Phi_1}-\frac{2ic_5^2C_1^2P}
{\sqrt{b}c_6\Phi_1\Phi_2}+\frac{c_6^2\Phi_2\Phi_1^3}{\sqrt{b}c_5^2C_1^2}
+\frac{ic_5(1+2C_1+2iabc_4C_2-2a)\Phi_1}{2\sqrt{b}a\Phi_2}.
\end{equation}
\end{subequations}

\bf Case 2 $c_5=0$. \rm In this case, $c_4=0$ should be hold at the same time because of the last component of \eqref{symp}. Without loss of generality, we can take $c_2=0,\ c_3=1$.
The final group invariant solution can be written as,
\begin{subequations}\label{red2}
\begin{equation}
\phi=\frac{c_1a}{c_6}\big[\tanh\big(F(\eta)+ac_1t\big)-1\big],\ \eta=x-ct,
\end{equation}
\begin{equation}
\phi_1=\Phi_1(\eta)\mbox{sech}\big(F(\eta)+ac_1t\big)
\exp\left(-\frac{c_1t}{2}\right),
\end{equation}
\begin{equation}
\phi_2=\Phi_2(\eta)\mbox{sech}\big(F(\eta)+ac_1t\big)
\exp\left(-\frac{c_1t}{2}\right),
\end{equation}
\begin{equation}
p=\left[P(\eta)-\frac{c_6}{c_1a}\Phi^2_1(\eta)
\tanh\big(F(\eta)+ac_1t\big)\right]\exp(c_1t),
\end{equation}
\begin{equation}
q=\left[Q(\eta)-\frac{c_6}{c_1a}\Phi^2_2(\eta)
\tanh\big(F(\eta)+ac_1t\big)\right]\exp(-c_1t).
\end{equation}
\end{subequations}
 The group invariant functions $F(\eta),\ P(\eta),\ Q(\eta),\ \Phi_1(\eta)$ and $\Phi_2(\eta)$ satisfy the following reduction equations
\begin{subequations}\label{redeq2}
\begin{equation}
F_\eta=\frac{c_6\Phi_1\Phi_2}{c_1a\sqrt{b}},
\end{equation}
\begin{equation}
Q=\frac{2i(c-2b\lambda)\Phi_2}{\sqrt{b}\Phi_1}+\frac{P\Phi_2^2}{\Phi_1^2}
-\frac{2ia^2c_1^2}{c_6\Phi_1^2},
\end{equation}
\begin{equation}
\Phi_{1\eta}=\frac{P\Phi_2}{\sqrt{b}}-i\lambda\Phi_1,
\end{equation}
\begin{equation}
\Phi_{2\eta}=\frac{Q\Phi_1}{\sqrt{b}}+i\lambda\Phi_2,
\end{equation}
\begin{equation}
P_\eta=\frac{P^2\Phi_2}{\sqrt{b}\Phi_1}-2iP\left(\lambda+\frac{c_1^2a^2}
{\sqrt{b}c_6\Phi_1\Phi_2}\right)+\frac{c_6^2\Phi_2\Phi_1^3}{\sqrt{b}
a^2c_1^2}
+\frac{(ic_1-2c\lambda+2b\lambda^2)\Phi_1}{\sqrt{b}\Phi_2}.
\end{equation}
\end{subequations}
The group invariant solutions \eqref{red1} and \eqref{red2}
hint us that we may develop a simpler while effective method to solve the AKNS system as for other nonlinear systems such as the KdV and KP equations \cite{n2}. It is noted that the exact solutions of the reduction systems \eqref{redeq1} and \eqref{redeq2} are left to be discussed at the end of the next section.

\section{Generalized tanh function expansion method}
As mentioned in the last section, the AKNS (and then NLS) system can be solved via a function expansion method, the generalized tanh function expansion method. Since the calculations are very simple, it is preferred to directly write down the following nonauto-B\"acklund transformation (NABT) theorem:\\
\bf NABT theorem. \rm If $w$ is a solution of
\begin{equation}\label{w}
\left(\frac{w_t}{w_x}\right)_t=\left(\frac{3w_t^2}{2w_x^2}+\frac12w_x^2
-\frac14\{w;\ x\}+4\lambda b \frac{w_t}{w_x}\right)_x, \quad \{w;\ x\}\equiv \frac{w_{xxx}}{w_x}-\frac32\frac{w_{xx}^2}{w_x^2},
\end{equation}
then
\begin{subequations}\label{pqw}
\begin{equation}
p=\sqrt{b} \left[w_x \tanh (w) -ib\frac{w_t}{w_x}-2i\lambda -\frac12\frac{w_{xx}}{w_x}\right]e^{iu},
\end{equation}
\begin{equation}
q=\sqrt{b} \left[w_x \tanh (w) +ib\frac{w_t}{w_x}+2i\lambda -\frac12\frac{w_{xx}}{w_x}\right]e^{-iu},
\end{equation}
\end{subequations}
is a solution of the AKNS system \eqref{AKNS} with the consistent conditions for the `phase' $u$
\begin{subequations}\label{uw}
\begin{equation}\label{ux}
u_x=2b\frac{w_t}{w_x}+2\lambda,
\end{equation}
\begin{equation}\label{ut}
u_t=3b\frac{w_t^2}{w_x^2}+8\lambda \frac{w_t}{w_x}+bw_x^2+6b\lambda^2-\frac b2\{w;\ x\}.
\end{equation}
\end{subequations}
The consistent condition $u_{xt}=u_{tx}$ of \eqref{uw} is nothing but \eqref{w}.

The theorem can be proved via two very simple means, the truncated Painlev\'e analysis and the tanh function expansion method. The Painlev\'e
analysis is one of the best approaches to study and solve special solutions for nonlinear physical systems. The function (especially the tanh function) expansion method is usually used to find traveling wave solutions of the nonlinear partial differential equations. Actually, the latter can be considered as the special case of the truncated Painlev\'e expansion approach.
Using the standard Painlev\'e analysis, it is straightforward to find the following auto- and nonauto-B\"acklund transformation (ANBT) theorem: \\
\bf ANBT theorem. \rm If $\phi$ is a solution of
\begin{equation}\label{phi}
\left(\frac{\phi_t}{\phi_x}\right)_t=\left(\frac{3\phi_t^2}{2\phi_x^2}
-\frac14\{\phi;\ x\}+4\lambda b \frac{\phi_t}{\phi_x}\right)_x,
\end{equation}
then both
\begin{subequations}\label{pqphi1}
\begin{equation}
p'=-\sqrt{b} \left[ib\frac{\phi_t}{\phi_x}+2i\lambda +\frac12\frac{\phi_{xx}}{\phi_x}\right]e^{iu},
\end{equation}
\begin{equation}
q'=\sqrt{b} \left[ib\frac{\phi_t}{\phi_x}+2i\lambda -\frac12\frac{\phi_{xx}}{\phi_x}\right]e^{-iu},
\end{equation}
\end{subequations}
and
\begin{subequations}\label{pqphi}
\begin{equation}
p''=\sqrt{b}\frac{\phi_x}\phi e^{iu}+p',
\end{equation}
\begin{equation}
q''=\sqrt{b}\frac{\phi_x}\phi e^{-iu}+q',
\end{equation}
\end{subequations}
are solutions of the AKNS system \eqref{AKNS} with the consistent conditions for $u$
\begin{subequations}\label{uphi}
\begin{equation}\label{ux1}
u_x=2b\frac{\phi_t}{\phi_x}+2\lambda,
\end{equation}
\begin{equation}\label{ut1}
u_t=3b\frac{\phi_t^2}{\phi_x^2}+8\lambda \frac{\phi_t}{\phi_x}+6b\lambda^2-\frac b2\{\phi;\ x\}.
\end{equation}
\end{subequations}
It is remarkable that \eqref{pqphi1} defines a NABT between the Schwarzian AKNS \eqref{phi} and AKNS \eqref{AKNS} while
\eqref{pqphi} defines an ABT between two solutions $\{p',q'\}$ and $\{p'',q''\}$ of the AKNS system. The consistent condition, $u_{xt}=u_{tx}$, of \eqref{uphi} is nothing but the Schwarzian AKNS \eqref{phi}.

Because both the quantities $\phi_t/\phi_x$ and $\{\phi;\ x\}$ are M\"obious transformation
\begin{equation}\label{MT}
\phi\rightarrow \frac{a\phi+b}{c\phi+d}, (ad-bc\neq 0)
\end{equation}
invariants, the Schwarzian AKNS (and NLS) equation \eqref{phi} is invariant under the M\"obious transformation (MT) \eqref{MT}. Actually, \eqref{MT} is just a special MT for $a=d=1,\ b=0,\ c=-\epsilon$ with its infinitesimal form \eqref{sf} because $\phi$ defined by \eqref{fxt} solves the Schwarzian AKNS \eqref{phi}.

Using the ANBT theorem and the following straightening transformation,
\begin{equation}\label{fw}
\phi=\frac{2}{\tanh(w)-1},
\end{equation}
the NABT theorem is naturally proved.

We call the transformation \eqref{fw} as the straightening transformation, because it transforms the single soliton solution of the Schwarzian AKNS \eqref{phi} and then the AKNS \eqref{AKNS} to a straight line solution $w=k_0x+\omega_0 t$. Furthermore, to add a soliton on any AKNS seed solution becomes a simple work to add a straight-line on $w$ equation, i.e., $w= k_0x+\omega_0 t+w_{seed}(x,t)$ for $w$ equation.
It is obvious that the solution expression \eqref{pqw} is the generalization of the usual tanh function expansion method. Actually, the transformation \eqref{fw} converts the usual truncated Painlev\'e expansion approach to the most general extension of the special tanh function expansion method.

In fact, the group invariant solutions \eqref{red1} and \eqref{red2} with \eqref{redeq1} and \eqref{redeq2} are just the special
cases of the theorem by taking
$$w=C_1\ln \tau +F(\xi),$$
and
$$w=ac_1t+F(\eta),$$
respectively.

Though the special function expansion and the truncated Painlev\'e
expansion have been utilized by many authors, no one can use them to generate interaction solutions between solitons and other types of waves such as the Painlev\'e waves, the cnoidal periodic waves and the quasi-periodic waves (algebra-geometric solutions).
All the known results obtained via the function expansion method, the truncated Painlev\'e expansion approach and other similar ones like the homogeneous balance method, the Riccati equation method and other auxiliary equation methods lose their generality. Thus, only some very special solutions such as the traveling or equivalent (which can be changed to) traveling waves (in new space-time) are obtained. Nonetheless, for instance, the NABT theorem is a general transformation without loss of any generality. Therefore, the method developed here can readily produce various new exact solutions (including those in the last section), which have not yet been found before. In the next section, we only focus on the interacting wave solutions between the soliton and cnoidal waves.

\section{Wave interaction solutions of the AKNS and NLS system}

Now, we use the theorem in the last section to find the interacting wave solutions of Eq. \eqref{w} with respect to $w$ or equivalently Eq. \eqref{phi} regarding $\phi$. The theorem shows that the single soliton (or solitary wave) solution of the AKNS system \eqref{AKNS} is only a straightened line solution $w=kx+\omega t$ of Eq. \eqref{w}. This fact reveals that to find the interaction solutions between solitons and other nonlinear excitations, we only need to find solutions in the form of
$$w=k_0x+\omega_0 t+v,$$
where $v$ is a function of $x$ and $t$ and satisfies
\begin{eqnarray}
&&v_{xx}(k_0+v_x)^4-\left(v_{tt}-4b\lambda v_{xt}+\frac14v_{xxxx}\right)(k_0+v_x)^2+\left[4(v_{xt}-b\lambda v_{xx})(\omega_0+v_t)+v_{xx}v_{xxx}\right](k_0+v_x)\nonumber\\
&&-3v_{xx}(\omega_0+v_t)^2-\frac34v_{xx}^3=0.\label{eqv}
\end{eqnarray}

Here, we just write down the soliton-cnoidal wave interaction solution for $\{w,\ u\}$ equations \eqref{w} and \eqref{uw},
\begin{subequations}\label{SC}
\begin{equation}\label{SCw}
w=k_0x+\omega_0t+c\ E_{\pi}(\mbox{sn}(kx+\omega t,m),n+1,m),
\end{equation}
\begin{eqnarray}\label{SCu}
u&=&u_0+\left(2\lambda+2b\frac{\omega_0}{k_0}\right)x
+b\left(\frac{m^2ck^3}{k_0(n+1)}+k_0^2+{6\lambda^2}
+8\lambda\frac{\omega_0}{k_0}+3\frac{\omega_0^2}{k_0^2}\right)t\nonumber\\
&&+\frac{2bc(k\omega_0-k_0\omega)}{kk_0(k_0+kc)} E_{\pi}\left(\mbox{sn}(kx+\omega t,m),\frac{k_0(n+1)}{kc+k_0},m\right),
\end{eqnarray}
\end{subequations}
where $\{u_0,\ m,\ n,\ k_1,\ k_2\}$ are arbitrary constants, other two arbitrary constants related to the initial center positions of the soliton and the cnoidal wave have been removed, and
\begin{eqnarray}\label{CC}
c^2&=&n-\frac{nm^2}{1+n},\quad \delta^2=1,\nonumber\\
\omega &=&-2bk\lambda-\frac{bk\delta\left[c^2k^3(c^2-n)-2ck_0k^2(n+nc^2-n^2-2c^2)
-3kk_0^2(2nc^2-c^2-n^2)-4nck_0^3\right]}
{2\sqrt{nck_0(kc+k_0)(kc-nk_0)(ck_0+kc^2-kn)}},\nonumber\\
\omega_0&=&-2bk_0\lambda+\frac{b\delta k_0
\left[c^2k^3(c^2-n)+kk_0^2(2nc^2-c^2-n^2)+2nck_0^3\right]}
{2\sqrt{nck_0(kc+k_0)(kc-nk_0)(ck_0+kc^2-kn)}}.
\end{eqnarray}
In the solution \eqref{SC}, $\mbox{sn}(z,m)$ is the usual Jacobi elliptic sine function and $E_{\pi}(\zeta,\ n,\ m)$ is the third type of incomplete elliptic integral defined by
$$E_{\pi}(\zeta,\ n,\ m)
       = \int_0^\zeta \frac{{\mbox{d}}t}{(1- n t^2)\sqrt{(1-t^2)(1-m^2t^2)}}.$$
Correspondingly, the physical quantity, the strength of the AKNS fields $\{p,\ q\}$, $I\equiv pq$ (i.e., $I=|p|^2$ for NLS), reads
\begin{eqnarray}\label{I}
I&=&b\frac{[kc-k_0(S^2-1)]^2}{(1-S^2)^2}\tanh^2(w)-\frac{2bck^2 S C D}
{(1-S^2)^2}\tanh(w)+4b\lambda^2+2\left(\frac{\omega_0}{k_0}
+\frac{\omega}{k}\right)\lambda+\frac{bk^2c^2}{(1-S^2)^2}\nonumber\\
&&+b\frac{\omega\omega_1}{kk_0}
+b\frac{c^2k^3(c^2-n)+kk_0(2nc^2-c^2-n^2)+2nck_0^3}{2nck_0}
+bk\frac{kc^2(2n-1)+n(2ck_0-kn)}{n(S^2-1)},
\end{eqnarray}
where $w$ is given by \eqref{SC}, $S\equiv \sqrt{n+1}\mbox{sn}(kx+\omega t,m),\ C\equiv \sqrt{n+1}\mbox{cn}(kx+\omega t,m), $ and $ D\equiv\mbox{dn}(kx+\omega t,m)$.
The soliton-cnoidal wave interaction structure expressed by \eqref{I} with
\eqref{SC} is abundant.  In the following, four types of special soliton-cnoidal waves are presented to show a variety of properties of the dark (gray) solitons under the background full of bright periodic cnoidal waves.

Fig. 1a exhibits the first type of special soliton-cnoidal wave structure of $I$ determined by Eq. \eqref{I} for the NLS equation at $t=0$ with the parameters selected as
\begin{equation}\label{cca}
\{b, k_1, k_2, \delta, \lambda, m, n, c, \omega_0, \omega\}=\{1, 1, -1, -1, 0, 0.999, -0.01, 0.00898989, -0.0919418, 0.202157\},
\end{equation}
while Fig. 1b displays its time evolution. From Fig. 1b, some interesting new important physics phenomena can be observed. (i) It is known that, in principle, the interactions between solitons are elastic, the only exchange is the `phase' shift (the shift of the soliton centers). Fig. 1b shows that the interaction between soliton and cnoidal wave (everyone peak of the cnoidal wave) is elastic except for a phase shift. (ii) Because the phase (the center of the soliton) will be shifted by every peak of the cnoidal periodic wave, the cumulative effect of the repeated phase shifts is equivalent to the deceleration and/or acceleration of the soliton. In the ocean, there are some
typical nonlinear waves which can travel in a very long distance. For instance, the tsunami waves in 2004 traveled 8000 km from epicenter (near Banda Aceh, Northern Sumatra,
Indonesia) to Port Elizabeth, South Africa! That means the tsunami waves have to be described by solitons or solitary waves. To describe the oceanic waves, the NLS equation \eqref{AKNS} with $q=p^*$ possesses the space variable $x$ and time $t$. Thus, if this NLS soliton is utilized to describe the tsunami waves, then Fig. 1b demonstrates the deceleration effect resulted from the interactions between the soliton and the background waves. It should be mentioned that the tsunami waves may be described by other soliton equations such as the Korteweg de Vries (KdV) and Kadomtsev-Petviashvili (KP) equations. The similar soliton-cnoidal wave interaction solutions and the same conclusions can be obtained \cite{n2,n1}.
In optics, especially for the fiber solitons, the NLS equation possesses the space variable $t$ and the time variable $x$. In this case, Fig. 1b reveals the acceleration effects due to the soliton-cnoidal wave interaction.

\input epsf
\begin{figure}[tbh]
\begin{center}
\epsfxsize=10.0cm\epsfysize=3cm\epsfbox{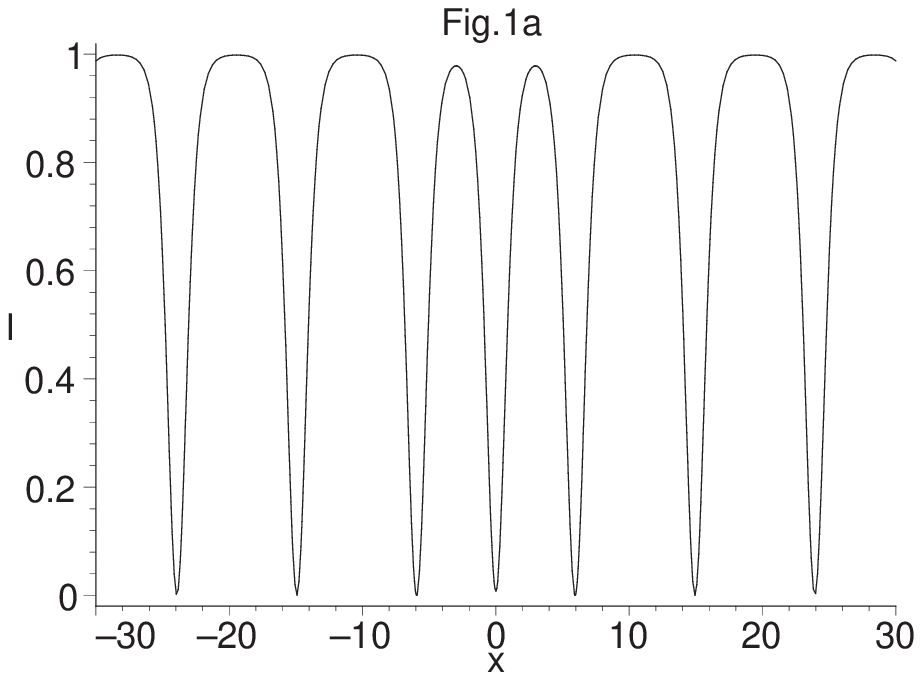}
\epsfxsize=10.0cm\epsfysize=8cm\epsfbox{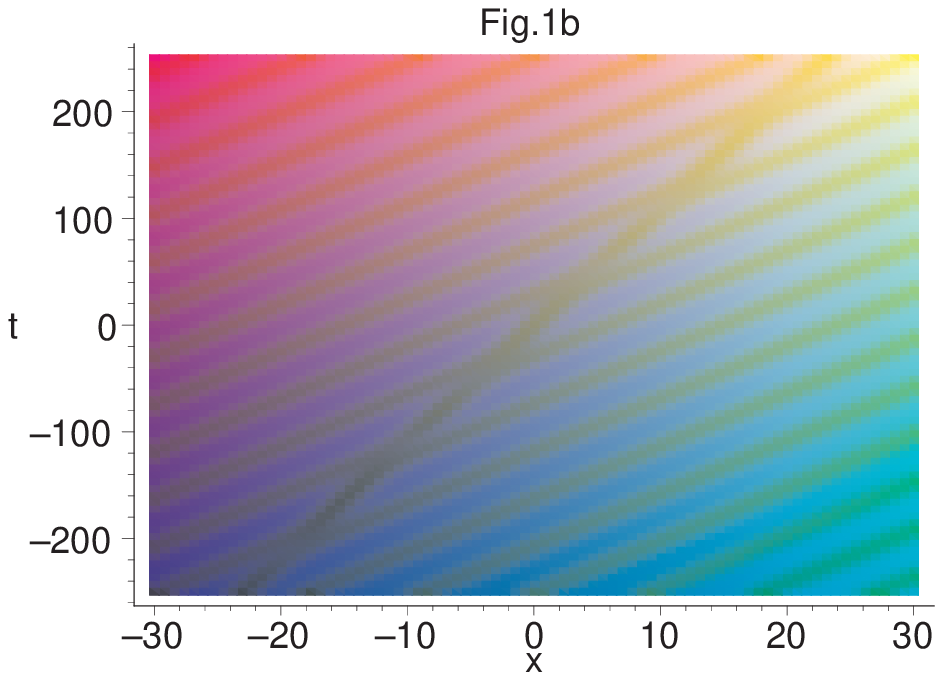}
\center{\footnotesize Fig 1: The first type of special soliton-cnoidal wave interaction solution for the NLS system given by \eqref{I} with the parameter selections \eqref{cca}: (a) The special structure at $t=0$; (b) The density plot for time evolution.}
\end{center}
\end{figure}
Fig. 2a displays the second type of soliton-cnoidal wave structure with the parameters fixed as
\begin{equation}\label{ccb}
\{b, k_1, k_2, \delta, \lambda, m, n, c, \omega_0, \omega\}=\{1, 1, 1, -1, 0, 0.9, -0.2, -0.05, 0.779897, 2.2540497\},
\end{equation}
at the initial time, and its time evolution is explored in Fig. 2b. It is seen from Fig. 2a that for small $m$ (not very close to $1$, here $m=0.9$) the cnoidal wave is different from the soliton lattice. However, the interaction between the soliton and every peak of the cnoidal periodic wave is also elastic with nonzero phase shifts.
\input epsf
\begin{figure}[tbh]
\begin{center}
\epsfxsize=10.0cm\epsfysize=3cm\epsfbox{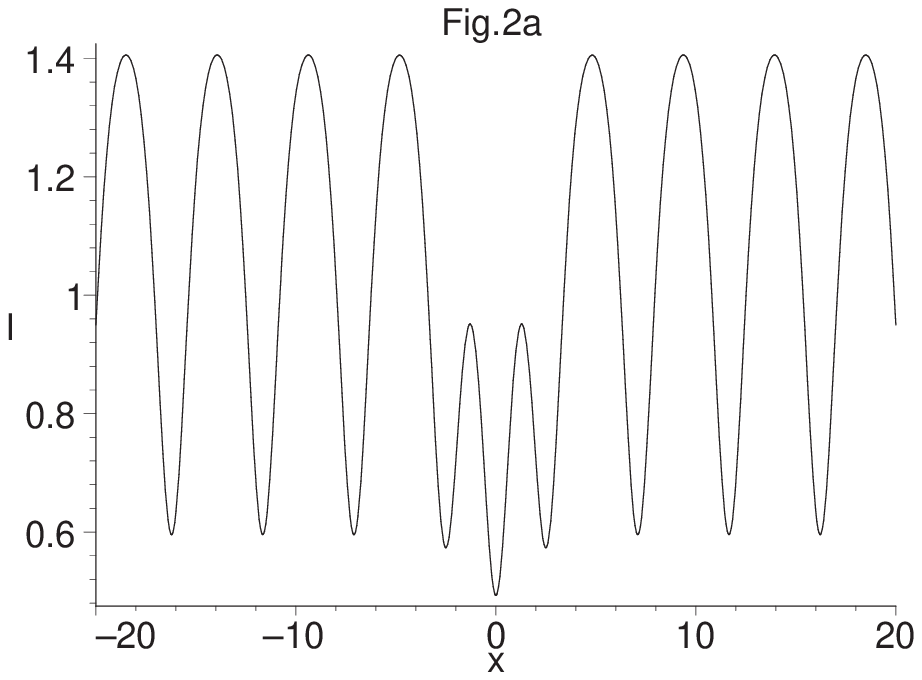}
\epsfxsize=10.0cm\epsfysize=8cm\epsfbox{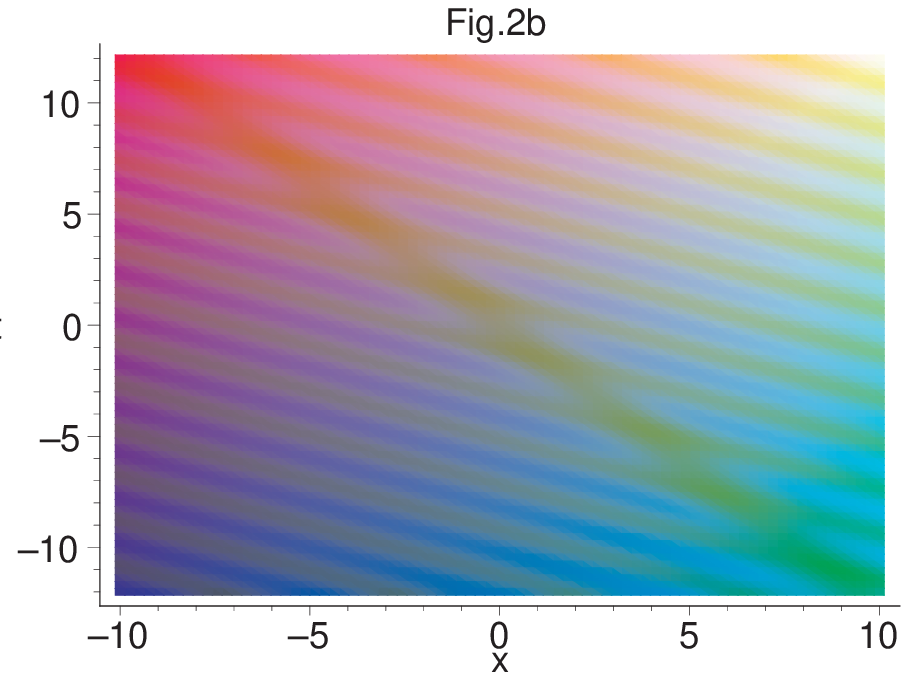}
\center{\footnotesize Fig. 2: The second type of special soliton-cnoidal wave interaction structure of \eqref{I} with the parameter selections \eqref{ccb}: (a) The special structure at $t=0$; (b) The density plot for time evolution.}
\end{center}
\end{figure}
Fig. 3a,\ Fig. 3b and Fig. 3c are the time evolutional plots for the third type of soliton-cnoidal wave interaction structure with the parameters determined as,
\begin{equation}\label{ccc}
\{b, k_1, k_2, \delta, \lambda, m, n, c, \omega_0, \omega\}=\{1, 1, 1, -1, 0.5, 0.999, -0.875, -2.4720451, -3.0730997, -0.858140\},
\end{equation}
at times $t=-10,\ 0$, and $10$, respectively. It is noted that if the parameter $0<n+1<1$ is fixed smaller and smaller in Figs. 1-3, then the relative depth of the gray soliton lattice is shallower and shallower compared with that of the single non-lattice soliton.

\input epsf
\begin{figure}[tbh]
\begin{center}
\epsfxsize=10.0cm\epsfysize=3cm\epsfbox{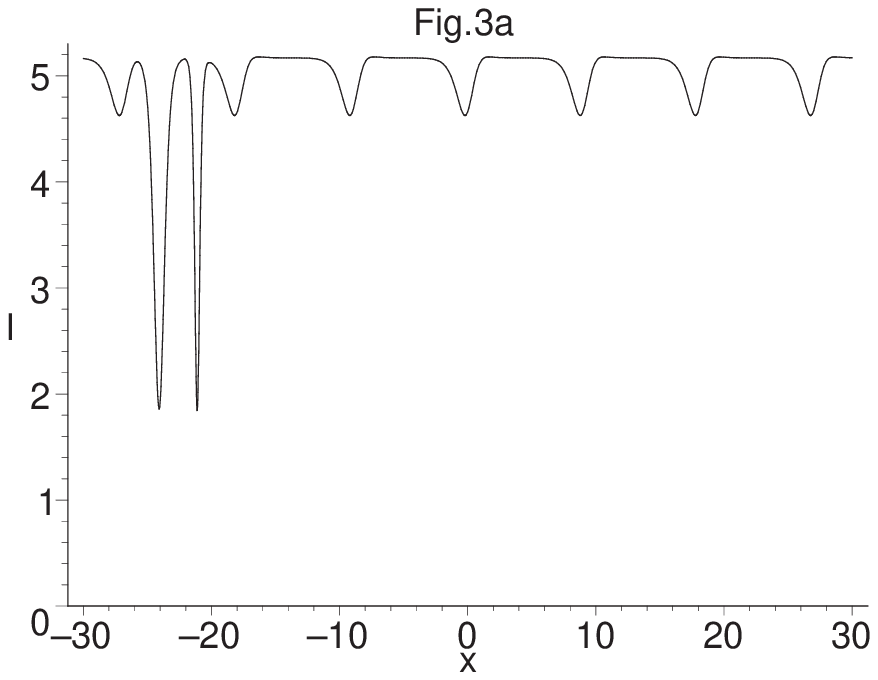}
\epsfxsize=10.0cm\epsfysize=3cm\epsfbox{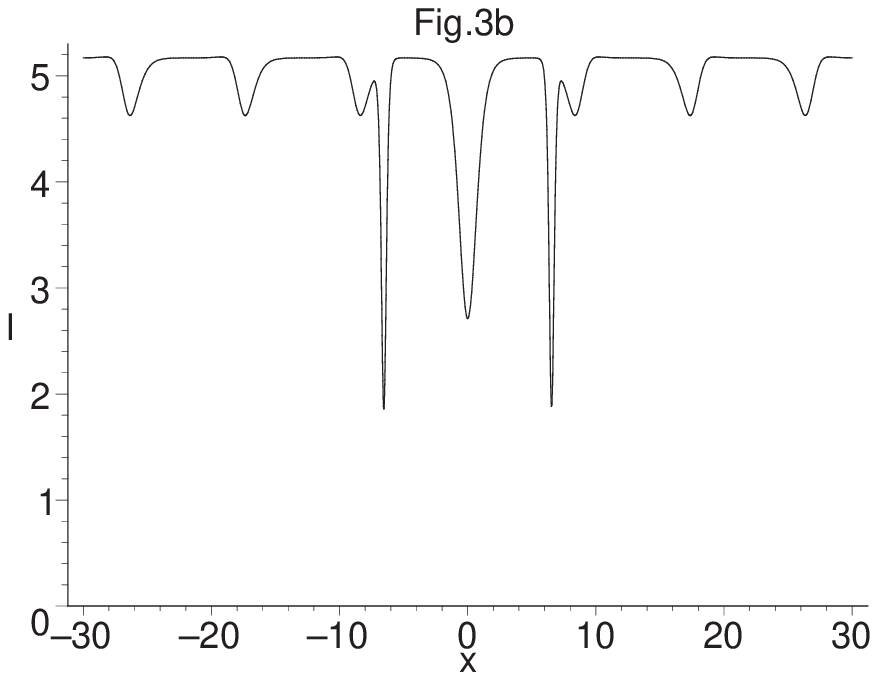}
\epsfxsize=10.0cm\epsfysize=3cm\epsfbox{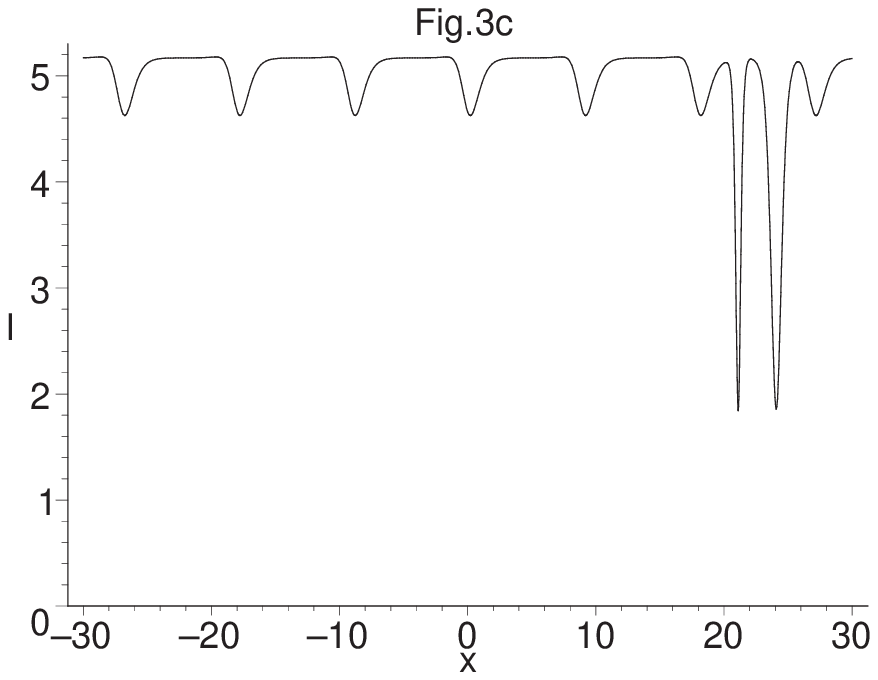}
\center{\footnotesize Fig. 3: Evolution of the third type of special soliton-cnoidal structure of the NLS system with the parameter selections \eqref{ccc} at (a) $t=-10,$ (b) $t=0$, and $t=10$, respectively.}
\end{center}
\end{figure}

Fig. 4 shows the fourth type of special soliton-conoidal wave interaction solution with the parameters being set as
\begin{equation}
 \label{ccc}
\{b, k_1, k_2, \delta, \lambda, m, n, c, \omega_0, \omega\}=\{1, 1, 1, -1, 0.5, 0.999, -0.00273650, -0.00142257, -0.665981, 0\}.
\end{equation}
It is clear that the optical soliton lattice (cnoidal wave) can also be static. In Fig. 4b, the moving soliton elastically interacts with every static lattice with the same phase shifts. From Fig. 4b, one can clearly get a novel interesting physical idea. Owing to the sudden shifts at the centers of the optical beams, the soliton-cnoidal wave interaction solution can be used to produce the controllable routing switches applicable in optical information and communication.

\input epsf
\begin{figure}[tbh]
\begin{center}
\epsfxsize=10.0cm\epsfysize=3cm\epsfbox{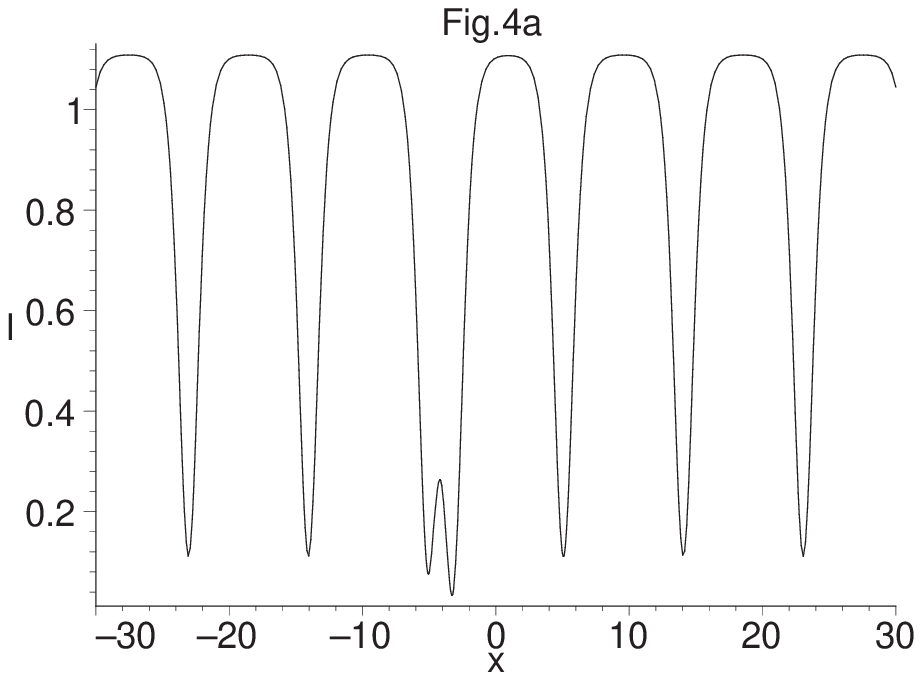}
\epsfxsize=10.0cm\epsfysize=8cm\epsfbox{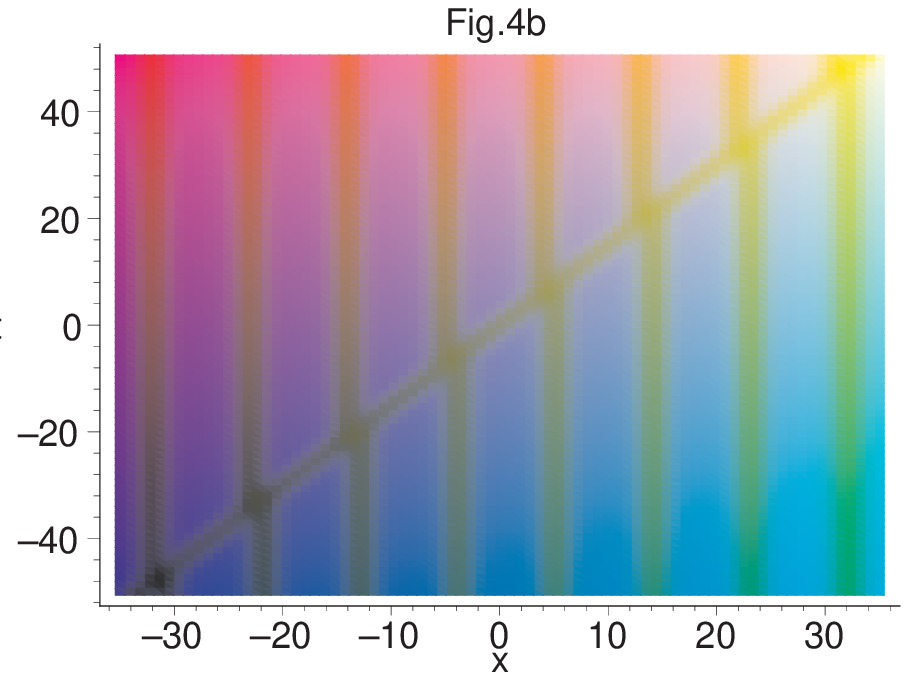}
\center{\footnotesize Fig. 4. The interaction solution of the NLS system for moving soliton and static lattice: (a) A plot of the special structure of \eqref{I} with \eqref{ccc} at $t=-5$; (b) Time evolution of (a).}
\end{center}
\end{figure}

It is not surprising that when $m\rightarrow 1$ in Eq. \eqref{SC}, the soliton-cnoidal wave interaction solution reduces to the two-soliton solution, whose interaction behaviors are displayed in Fig. 5 with the parameters selected as
\begin{equation}\label{ccd}
\{b, k_1, k_2, \delta, \lambda, m, n, c, \omega_0, \omega\}=\{1, 1, 1, -1, 0, 1, -0.95, -4.248259, -5.894616, 0.3634499\}.
\end{equation}

\input epsf
\begin{figure}[tbh]
\begin{center}
\epsfxsize=10.0cm\epsfysize=8cm\epsfbox{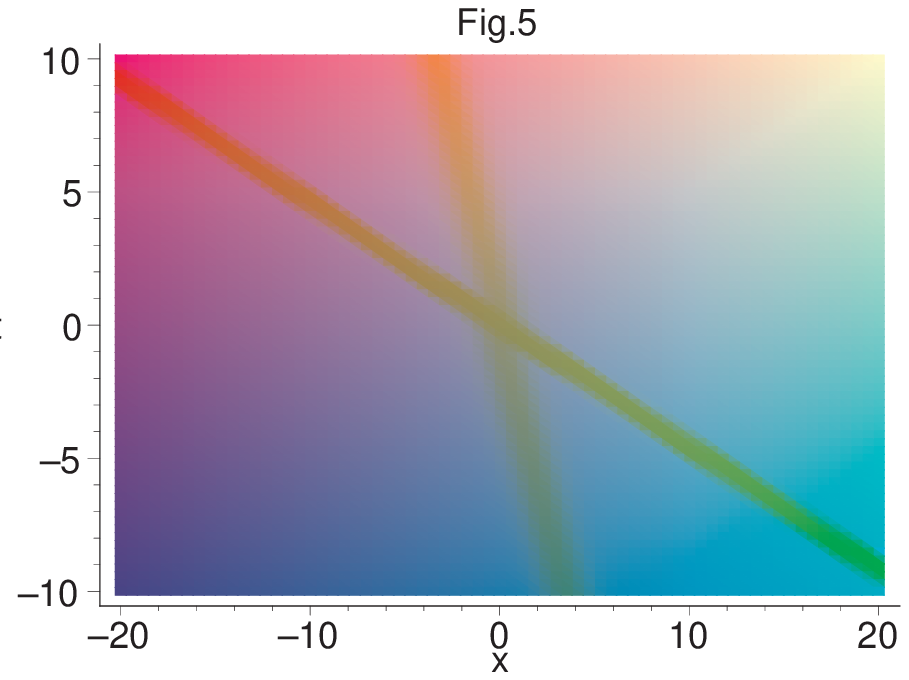}
\center{\footnotesize Fig. 5. The special case of soliton-cnoidal wave interaction, two soliton case \eqref{I} with \eqref{ccd}.}
\end{center}
\end{figure}

\section{Summary and discussions}
To sum up, infinitely many nonlocal symmetries of the AKNS system are obtained by expanding the square eigenfunction symmetry with respect to the spectral parameter $\lambda$ or equivalently by means of infinitely many Lax pair \eqref{Lxn} and \eqref{Ltn}. Though the square eigenfunction symmetry \eqref{s0} is nonlocal for the original AKNS system, it is localized for the enlarged system \eqref{AKNS}, \eqref{Lx}, \eqref{Lt} and \eqref{fxt}. The prolonged AKNS system is studied by symmetry reductions with localized symmetries which is nonlocal for the original system. Hinted by the results of the symmetry reductions, two quite simple but equivalent methods, the truncated Painlev\'e expansion approach and the generalized tanh function expansion method, are established to explore interactions between different nonlinear excitations. Especially, the soliton-cnoidal wave interaction solutions are explicitly expressed by the Jacobi elliptic functions and the third type of incomplete elliptic integral. In order to extend the known method to discover new phenomena in nonlinear physics, the key point is the straightening transformation \eqref{MT} which straightens the single soliton to a straight line solution and transforms the truncated Painlev\'e expansion to an extended tanh function expansion method.
The methods are valid for all integrable systems and even nonintegrable models because both the truncated Painlev\'e analysis and the tanh expansion method can generate exact solutions of partially solvable nonlinear models. The theorems can be used to find interaction solutions between solitons and any other NLS waves. However, for simplicity, only the soliton-cnoidal wave interaction solutions are discussed in detail.

The soliton-cnoidal wave interaction solutions display some interesting new physical phenomena. The first observation is that the interactions between the soliton and the background waves are elastic with phase shifts. The further consideration implies that the accumulated effect due to the interactions between soliton and the background waves is equivalent to the deceleration for the oceanic solitary waves (such as the tsunami waves) and the acceleration for the optical fiber solitons.
Another potential application of the soliton-cnoidal interaction solution is to offer a new mechanism to produce the controllable routing switches in optical information and communications.

The methods and what obtained in this paper are valid to all the integrable models. The details on the methods for other nonlinear systems, other types of interacting wave solutions, other methods to solve interaction solutions between different types of nonlinear excitations, other possible new physical applications, and so on, will be reported in our future research work.

The work was sponsored by the National Natural Science Foundations of
China (Nos. 11175092, 11275123, 11205092 and 10905038), Shanghai Knowledge Service Platform for Trustworthy Internet of Things (No. ZF1213), Scientific Research Fund of Zhejiang Provincial Education Department under Grant No. Y201017148,
and K. C. Wong Magna Fund in Ningbo University.

\small{
}
\end{document}